\documentclass[useAMS,usenatbib]{mn2e}
\usepackage{graphicx,color,epsfig}

\newcommand{\be}{\begin{equation}}
\newcommand{\ee}{\end{equation}}

\newcommand{\apj}{ApJ}

\newcommand{\mnras}{MNRAS}
\newcommand{\aap}{A\&A}

\newcommand{\nat}{Nature}

\def\ltsima{$\; \buildrel < \over \sim \;$}
\def\simlt{\lower.5ex\hbox{\ltsima}}
\def\gtsima{$\; \buildrel > \over \sim \;$}
\def\simgt{\lower.5ex\hbox{\gtsima}}

\def\msun{{\,{\rm M}_\odot}}

\newcommand\mearth{{\,{\rm M}_{\oplus}}}

\def\del#1{{}}

\title[Disrupted hot jupiters]{Hot Super Earths: disrupted young jupiters?}

\author[S. Nayakshin]{Sergei Nayakshin\\ 
Department of Physics \& Astronomy,
  University of Leicester, Leicester, LE1 7RH, UK\\
{E-mail:~} {\rm Sergei.Nayakshin@astro.le.ac.uk}}

\begin{document}

\date{Received}

\pagerange{\pageref{firstpage}--\pageref{lastpage}} \pubyear{2008}

\maketitle

\label{firstpage}

\begin{abstract}
Recent {\em Kepler} observations revealed an unexpected abundance of ``hot''
Earth-size to Neptune-size planets in the inner $0.02-0.2$ AU from their
parent stars. We propose that these smaller planets are the remnants of 
massive giant planets that migrated inward quicker than they could
contract. We show that such disruptions naturally occur in the framework of
the Tidal Downsizing hypothesis for planet formation. We find that the
characteristic planet-star separation at which such ``hot disruptions'' occur
is $R \approx 0.03-0.2$~AU.  This result is independent of the planet's embryo
mass but is dependent on the accretion rate in the disc. At high accretion
rates, $\dot M \simgt 10^{-6}\msun$~yr$^{-1}$, the embryo is unable to
contract quickly enough and is disrupted. At late times, when the accretion
rate drops to $\dot M \simlt 10^{-8} \msun$~yr$^{-1}$, the embryos migrate
sufficiently slow to not be disrupted. These ``late arrivals'' may explain the
well known population of hot jupiters. If type I migration regime is
inefficient, then our model predicts a pile-up of planets at $R\sim 0.1$ AU as
the migration rate suddenly switches from the type II to type I in that
region.
\end{abstract}


\section{Introduction}\label{sect:intro}

Standard proto-planetary disc models \citep{CG97} show that the inner $\sim
0.1$ AU region is too hot to allow existence of small solid particles
there. Thus planets should not be able to grow there. Yet observations of
nearby solar type stars show that many of them do host planets in that
inhospitable to planet formation region. The very first exoplanet to be
convincingly detected had the separation of $R \sim 0.05$ AU to its parent
star and had a mass of about that of Jupiter \citep{MQ95}. Such gas giant
planets circling their parent stars in a close proximity ($R\simlt 0.1$ AU)
are now called ``hot jupiters''. It is believed that they are explained by the
inward radial drift (migration) of the planets born further out \citep{Lin96}.

The {\em Kepler} mission has recently produced a number of surprising results
\citep{BoruckiEtal11}, one of which is that there is an even greater number of
smaller planets, e.g., Earth-size to Neptune-size, in that region. It is
similarly obvious that these smaller ``hot'' planets also had to be brought
there from further out by an inward radial migration. One practical difficulty
in testing this idea, though, is that the migration of smaller planets is
expected to occur in the poorly understood ``type I'' regime
\citep{PP08}. This is different from the better understood ``type II'' regime
by which the giant planets migrate \citep{LinPap86}. This theoretical
difficulty leads to a large range in uncertainties in the predictions of the
detailed Core Accretion model calculations \citep[e.g., see Fig. 5
  in][]{IdaLin08}.

Recently, the key importance of the radial migration of the earliest gas
condensations formed in the massive proto-planetary discs by the gravitational
instability was realised \citep{VB06,BoleyEtal10}. Analytical estimates
\citep{Nayakshin10c} and numerical simulations
\citep{VB06,VB10,BoleyEtal10,ChaNayakshin10} show that these condensations can
migrate all the way from their birth-place in the outer $R\sim 100$ AU disc
into the inner $\sim$~few AU disc and be disrupted there during the earliest
massive disc phase ($t\simlt$ few$\times 10^5$ yrs, typically).

It was pointed out by \cite{BoleyEtal10} and \cite{Nayakshin10c} that this
migration-and-disruption sequence yields an unexplored way of forming
terrestrial like planets. If dust grows and sediments in the centre of the
clump and forms a solid density core there, then tidal disruption of the gas
clump may leave a solid core -- an Earth-like proto-planets \citep[note the
  connection to earlier ideas
  of][]{McCreaWilliams65,Boss98,BossEtal02}. \cite{Nayakshin10a,Nayakshin10b}
used a simple spherically symmetric radiation hydrodynamic code with the dust
grains as a second fluid to delineate the conditions when such a mechanism for
the solid core growth can work. Based on the potential promise of these ideas,
\cite{Nayakshin10c} formulated the ``Tidal Downsizing'' \citep[TD
  hereafter;][]{Nayakshin10c} hypothesis for planet formation. In this picture
a partial disruption of a $\sim 10 M_J$ gas clump (which we also call giant
embryos; GEs) leaves a giant planet, whereas a complete disruption yields a
terrestrial like planet.

In this Letter we continue to assess the potential utility of the TD
hypothesis to planet formation.  We note that another ingredient, muted but
not explicitly considered by
\cite{BoleyEtal10,Nayakshin10a,Nayakshin10b,Nayakshin10c}, must be included in
the scheme. To explain it, consider isolated GEs first. As they contract,
their internal temperature increases. At early times the rate of this
contraction is controlled by the radiative cooling rate of the embryo -- which
is by the rate at which the embryo can get rid of the excess energy.  However,
when temperature $T_{\rm 2nd} \approx 2000$ K is reached, molecular hydrogen
disassociates. This process is an efficient energy sink, which allows the
embryo to contract rapidly -- in fact collapse hydrodynamically -- without the
need to radiate the energy away. The embryo collapse stops only at much higher
densities, and temperatures as high as $10^4$ K, at which point hydrogen is
ionised. The embryo must then continue a slower contraction, again regulated
by the rate at which its energy is radiated away. The collapse is known as the
``second collapse'' in the star formation literature \citep{Larson69}, when
the ``first cores'' of masses $\sim 50 M_J$ collapse \citep{Masunaga00} to
become ``second cores'', which are the proper proto-stars.

In the TD hypothesis for planet formation, the second collapse may be the last
step to making a gas giant planet. However, as we show below, this final step
is not automatically successful -- planets continuing to migrate rapidly
towards their parent stars may still be disrupted at $R \sim 0.1$ AU. We
suggest this process as a way of forming the hot Super Earths observed by the
{\em Kepler} mission \citep{BoruckiEtal11}. 

\section{The second embryo}

In analogy to the star formation literature, we refer to the GEs that are
mainly molecular, embryo's temperature $T_e < T_{\rm 2nd}$, as the ``first
GEs''; those where H$_2$ is disassociated are termed ``second GEs'' instead.

\subsection{Contraction and collapse of the first embryo}\label{sec:1st}

To illustrate the main point of this paper, we calculate the contraction of a
giant embryo with ``typical'' parameter values \citep[e.g., those that appear
  quite reasonable to us for a solar metalicity disc around a $\sim$ solar
  mass star; see][]{Nayakshin10c}. In particular, the embryo mass is $M_e = 10
M_J$, the normalised dust opacity is $k_* = 0.5$, and the grain mass fraction
$f_g = 0.01$. The embryo is initialised as a first core of same mass
\citep[see][]{Nayakshin10a}.

\begin{figure}
\centerline{\psfig{file=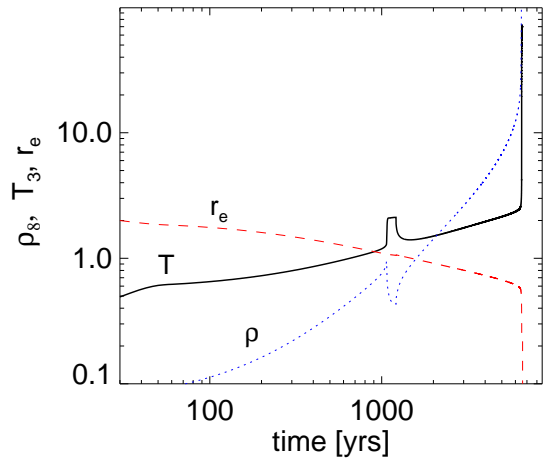,width=0.44\textwidth,angle=0}}
\caption{The embryo's temperature (in units of $10^3$~K), density (in
  $10^{-8}$ g cm$^{-3}$) and radial size, $r_e$, in AU, as a function of time,
  as labelled on the figure. Note the abrupt change near the end of the
  calculation, marking the second collapse, when the central temperature
  reaches $T\sim 2300$~K. The bump in the central temperature near $t=1000$
  yrs is caused by the formation of a $\approx 5\mearth$ solid core inside the
  embryo.}
\label{fig:fig1}
\end{figure}

Figure \ref{fig:fig1} shows the time evolution of the embryo's central
temperature (solid, in units of $10^3$~K), the gas density (dotted, in units
of $10^{-8}$ g cm$^{-3}$), and the outer radius of the embryo, $r_e$, (dashed,
in units of 1 AU). The calculation is carried out with an updated version of
the 1D gas-dust grains radiative hydrodynamics code of
\cite{Nayakshin10a,Nayakshin10b}. Instead of using an ideal gas equation of
state with $\gamma = 5/3$, the code now uses the equation of state appropriate
for molecular hydrogen, including disassociation and rotational and
vibrational degrees of freedom for H$_2$, with the orto-hydrogen to
para-hydrogen ratio fixed at 3:1 \citep[cf.][]{BoleyEtal06}.

Despite the updated equation of state, the evolution of the first embryo is
quite similar to that of the cases studied in \cite{Nayakshin10b}. This may
not be entirely surprising given the similar insensitivity of the first (gas)
cores to the equation of state as found by \cite{Masunaga98,Masunaga00}.  The
embryo contracts and heats up, whereas dust grains grow. By time $t\sim 1000$
yrs, the grains increase in size to about 20 cm. Their density exceeds that of
the gas in the centre of the embryo; they become self-gravitating and form a
solid core of mass $M_{\rm c} \sim 5\mearth$. In Figure \ref{fig:fig1}, the
solid core formation is notable by the bump in the central temperature. After
the core formation, the central region becomes hotter than grain vaporisation
temperature of $\approx 1400$~K, evaporating the grains, and thus terminating
further core growth \citep[see][for details on this negative feedback
  loop]{Nayakshin10b}. The central region also expands slightly. Most of the
GE is however unaffected by the solid core in this case, and the curves resume
their otherwise monotonic behaviour a few hundred years later.

At $t\approx 6.5\times 10^3$ yrs, the GE goes through the second collapse when
the central temperature exceeds about 2300 K. The embryo radius drops rapidly,
while the density and the temperature increase strongly. The first embryo
becomes the second in our terminology.

\subsection{The contraction of the second embryo}

Our radiation hydrodynamics code is not well suited \citep{Nayakshin10b} to
follow the long cooling and contraction of the second embryo. However, there
is a body of work on the hydrostatic contraction of very low-mass stars and
Jupiter-mass planets \citep[e.g.,][]{GrossmanGraboske73,GraboskeEtal75}, which
allows us to describe the process with a good degree of confidence. Let $t_2$
be the time of the second collapse of the embryo. As is well known, the embryo
spends the initial contraction stages on the Hayashi track. During this phase,
the outward energy transfer is dominated by convection. The effective
temperature, $T_{\rm eff}$ is almost constant at $\log T_{\rm eff} \sim
3.1-3.3$ [see Figs 1 in \cite{GrossmanGraboske73,GraboskeEtal75}]. The
luminosity of the embryo is then $4\pi r_e^2 \sigma T_{\rm eff}^4$.

With this simple cooling model, we solve for the evolution of the second GE
radius, $r_e(t)$,
\begin{equation}
{dE_{\rm GE}\over dt} =  4\pi r_e^2 \sigma_B T_{\rm eff}^4\;,
\end{equation}
where the energy of the GE is defined with the positive sign as $E_{\rm GE}
\approx GM_e^2/2r_e$. A trivial integration yields\footnote{A similar model
  can be used to describe the contraction of young $M_*\sim \msun$ proto-stars
  as well \citep[e.g.,][]{Cameron95}.},
\begin{equation}
r_e^3(t) = {r_{2}^3 \over 1 + A r_{2}^3 (t-t_2)}\;.
\label{re_t}
\end{equation}
Here $r_2$ is the GE radius at the time of the second collapse, $t_2$, and $A
= 24 \pi \sigma_B T_{\rm eff}^4/(GM_e^2)$. This model does not take into
account the electron degeneracy pressure in the GE, but that becomes important
only after $t\simgt 10^6$ yrs, and would make disruption of the planets even
more likely.

Evaluation of equation (\ref{re_t}) shows that at times $(t-t_2) \simgt 100$
yrs, the initial radius of the embryo is quickly forgotten, and the
contraction proceeds as
\begin{equation}
r_e(t) = \left({1 \over A (t-t_2)}\right)^{1/3}\;.
\label{re_ass}
\end{equation}
The second embryo's density as a function of time is
\begin{equation}
\rho_e(t) = {3M_e A (t-t_2) \over 4\pi} = {18 \sigma T_{\rm eff}^4  \over
  GM_e}\left(t-t_2\right) \;.
\label{rho_ass}
\end{equation}
In terms of absolute values, the initial density of the second embryo is at
least $10^{-6}$ g cm$^{-3}$ (cf. Fig. 1) and rises with time according to
\begin{equation}
\rho_e(t) = 6\times 10^{-4} \;\hbox{g cm}^{-3}\; {(t-t_2)\over 10^4\;\hbox{yr}}
    {10 M_J \over M_e}\;,
\label{rho2_est}
\end{equation}
where we set $\log T_{\rm eff} = 3.1$ for certainty.

\subsection{Vulnerability of young massive planets}\label{sec:young}

A distance $R$ away from the star, the tidal density is
\begin{equation}
\rho_{\rm t} = \frac{M_*}{2 \pi R^3} \; \approx 10^{-7} \;\hbox{g cm}^{-3}\;
\frac{M_*}{\msun}R_{AU}^{-3}\;,
\label{rhotd}
\end{equation}
where $M_*$ is the stellar mass, and $R_{AU}= R/$1 AU. A giant embryo that
migrated to a distance $R$ is disrupted if the embryo's density $\rho_e\simlt
\rho_{\rm t}$.  The first and the second embryos have considerably different
internal densities.  The ``first embryos'' are characteristically disrupted
at $R\simgt 2$ AU \citep{Nayakshin10c}. We term these disruptions ``cold'' and
do not consider any further here.

To evaluate the second GE's fate, we need to estimate its migration rate. We
assume that there is only one such GE in the inner $R \simlt $ few AU at any
one time. Due to its significant mass, the GE migrates in the type II
regime. As shown by \cite{IvanovEtal99}, the migration time, $t_{\rm mig}$, is
shorter than but is comparable to the ``accretion time'', $t_{\rm a} = M_e/\dot
M$, where $\dot M$ is the accretion rate in the disc. In the TD model, the
planets are born in the very early gas-rich phase of the disc when the
proto-star is far from its final mass.  The accretion rate during that phase
is $\dot M \sim M_*/t_{\rm ff}$, where $t_{\rm ff}\sim 10^5$ yrs is the
free-fall time for typical interstellar molecular gas cores of $\sim 1 \msun$
mass \citep{Larson69}. Thus,
\begin{equation}
t_{\rm mig} \simlt {M_e\over \dot M} = 10^3 \;\hbox{yrs}\; {M_e \over 10 M_J}\;
{10^{-5} \msun \hbox{yr}^{-1}\over \dot M}\;.
\label{tmigr}
\end{equation}
Now, this time scale is to be compared with the Kelvin-Helmholtz time scale
for isolated giant planets, which is of the order of $t_{\rm KH}\sim
(10^4-10^5)$ yrs \citep[e.g., see Figs. 2 \& 3 of][]{GraboskeEtal75}.  In
fact, any additional effects we can think off, e.g., stellar irradiation
\citep[e.g.,][]{CameronEtal82}, tidal heating, energy stored in the rotation
of the GEs, accretion of planetesimals, etc., should only increase $t_{\rm
  KH}$.

We can now estimate the embryo-star separation at which the disruption occurs,
$R_{\rm hot}$, e.g., where $\rho_e = \rho(R_{\rm hot}$. For this we note that
$(t-t_2)$ in equation (\ref{rho_ass}) should be of the order of $t_{\rm
  mig}$. Indeed, this is the time it takes the embryo to migrate inwards, and
thus the embryo's age should be comparable to that. Entering $t_{\rm mig}$
instead of $(t-t_2)$ in equation (\ref{rho_ass}), we find that the dependence
on the embryo's mass cancels out, and we arrive at
\begin{equation}
R_{\rm hot} = \left[{GM_*\dot M\over 36\pi \sigma T_{\rm eff}^4}\right]^{1/3}
= 0.12 \;\hbox{AU} \left({\dot M \over 10^{-5} \msun \hbox{yr}^{-1} }\right)^{1/3}\;,
\label{rhot_t}
\end{equation}
where we set $\log T_{\rm eff} = 3.1$ and retained the dependence on $\dot M$
only. We term this inner disruption ``hot'' to distinguish from the more
distant cold disruptions.  Equation \ref{rhot_t} shows that a second GE
entering the inner $\sim 0.1$ AU in an early disc phase, when $\dot M$ is very
high, is likely to be tidally disrupted. On the other hand, ``late arrivals'',
when $\dot M \simlt 10^{-8} \msun$ yr$^{-1}$, may survive and contract into
hot jupiters as the disc runs out of mass, presumably due to photo-evaporation
\citep{AlexanderEtal06}.

\section{An illustrative model}\label{sec:model}

We now present two approximate example calculations that are complete in the
sense that they start off with the embryo's birth at $R = 100$ AU and they end
with the embryo arriving in the innermost disc. We use the model embryo
calculated in \S 2.1 and 2.2, and an approximate radial migration model of
\cite{Nayakshin10c}. We consider two opposite limiting cases to illustrate the
points made in \ref{sec:young}.

\subsection{A disrupted hot jupiter}\label{sec:disrupt}

\begin{figure}
\centerline{\psfig{file=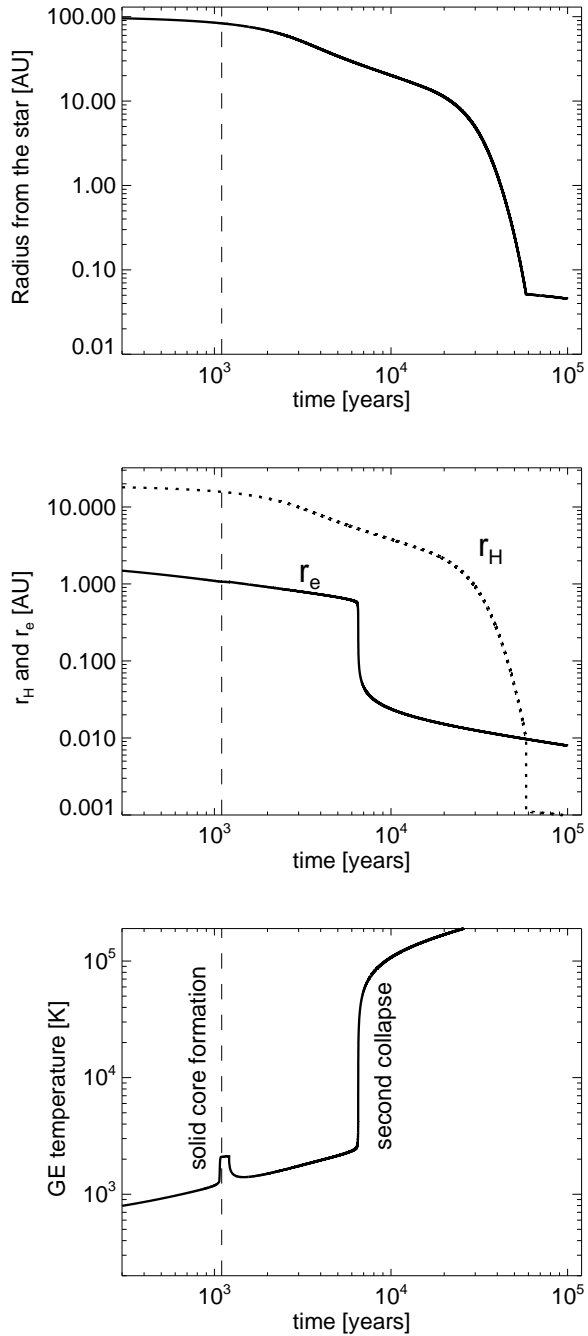,width=0.46\textwidth,angle=0}}
\caption{Formation of a ``hot Super Earth'' starting from a giant embryo of 10
  Jupiter masses. {\bf Upper panel}: Radial position of the embryo versus
  time. The GE starts off at $R=100$ AU. The vertical dashed line marks the
  time when a solid core forms inside the embryo. {\bf Middle}: Radius of the
  embryo, $r_e$, and the Hills radius, $r_H$. The GE gaseous envelope is
  assumed to be completely disrupted when the two sets of curves meet. {\bf
    Lower}: Temperature of the first cores as a function of time. Note the
  strong jump at the second collapse.}
\label{fig:fig2}
\end{figure}

In the first case, presented in Figure \ref{fig:fig2}, the accretion disc is
massive and the accretion rate is high. In particular, the initial mass of the
star is $M_* = 0.5\msun$, the doubling timescale for the star, $t_{\rm db} =
10^5$ yrs. This yields an accretion rate through the disc of $5\times
10^{-6}\msun$~yr$^{-1}$. At $t=t_{\rm db}$, the assembly of the star is
assumed complete (its mass reaches 1 $\msun$), and the disc torques are abruptly
removed.

For simplicity we assume that the mass of the GE is constant until it is
disrupted dynamically.  3D simulations of embryos migrating in gas discs show
destruction of the embryos in a matter of two or three orbits
\citep{BoleyEtal10,ChaNayakshin10} once they fill their Roche lobes.

The upper panel of Figure \ref{fig:fig2} shows the radial location of the
embryo. The embryo migrates into the inner $\sim 0.1$ AU region of the disc in
less than $10^5$ yrs due to the high disc accretion rate. The middle panel
shows the evolution of the Hills radius, $r_H$, and the embryo's radius,
$r_e$. $r_e$ is initially one to two orders of magnitude smaller than the
Hills radius, especially right after the second collapse. The lower panel
shows the embryo's temperature evolution, which is comprised of two parts: the
first one calculated as in \S 2.1 (and shown in Figure 1) until the second collapse
occurs, at which point we switch to the model of \S 2.2, with $T$ found
through the virial relation of the embryo.

The ``hot'' disruption of the embryo occurs at time $t\approx 6\times 10^4$
yrs, at the embryo-star separation of $R=0.06$ AU. Note that disruption of the
gaseous envelope leaves a solid core behind since the solid core formation
occured much earlier. The solid core migrates slightly to $R\sim 0.045$ AU by
$t=t_{\rm db}$. This last bit of radial migration is via type I and may in
reality be far less efficient \citep{PP08,IdaLin08}.

\subsection{A tidally stable hot jupiter}\label{sec:not_disrupt}

\begin{figure}
\centerline{\psfig{file=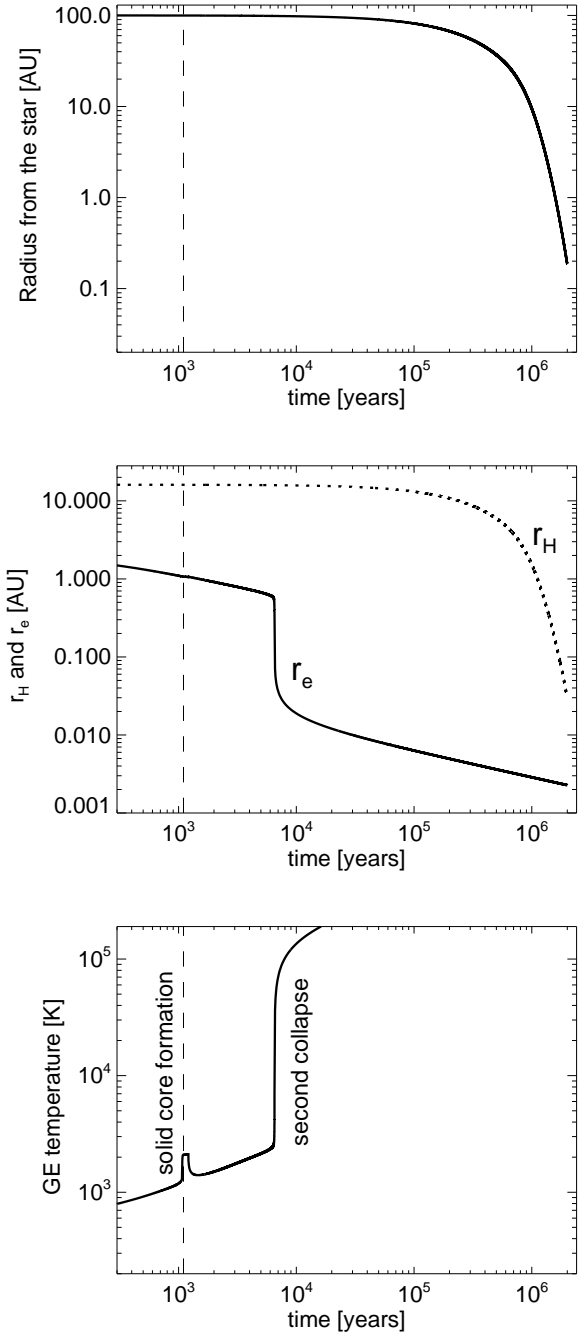,width=0.46\textwidth,angle=0}}
\caption{Same as Figure \ref{fig:fig2} but for the disc accretion rate of
  $10^{-7}\msun$ yr$^{-1}$. The embryo is not disrupted in the inner
  disc since it has had enough time to contract to much higher densities.}
\label{fig:fig3}
\end{figure}

In the other example we take a proto-star that is closer to its final mass,
$M_* = 0.8\msun$, accreting at a rate of $10^{-7}\msun$~yr$^{-1}$ for $2\times
10^6$ yrs. Figure \ref{fig:fig3} shows that in this case the embryo migrates
much slower. By the time it arrives in the innermost disc, the embryo radius,
$r_e$, dropped to about $2\times 10^{-3}$ AU. The embryo is too
compact to be disrupted unless it migrates even further to $\sim 0.01$ AU (it
is likely to stall before that due to the magnetospheric cavity in the inner
disc).

\section{Discussion}

We suggested here that massive gaseous proto-planets in the ``second''
configuration, migrating rapidly inward, are tidally disrupted in the inner
$\sim 0.1$ AU. If these protoplanets contain solid cores, then the cores
remain unaffected by the disruption as their densities are much higher. For
solid cores more massive than $\sim 20\mearth$, parts of the gaseous envelopes
may be retained as well \citep{Nayakshin10b}. Therefore, the hot disruption of
gas giant planets could in principle result in the production of purely solid
and also solid plus gas envelope planets similar to those found in the {\em
  Kepler} data. Further modelling of the regions closest to the solid cores
inside the embryos is needed for a more quantitative statement.


The outcome of a hot jupiter arriving in the inner $0.1$ AU strongly depends
on the migration rate of the latter and its age. If the proto-planet is
``old'', e.g., $\simgt 1$ Myrs, then it is too dense to be disrupted near the
star.  The observed hot jupiters in our
interpretation are the planets that arrived in the inner $0.1$ AU somewhat
late, when they were already compact, and when the disc was running out of
material.


Some of the {\em Kepler} Neptune and Earth-size planets could have parted with
their gaseous envelopes in the cold disruptions at a few AU, and then migrated
inward in the disc. This would make for a second channel of making these
planets in the inner disc. However, the location of hot disruption is quite
well defined, $R\sim 0.03 - 0.2$, due to the strong dependence of the tidal
density on $R$. Therefore we predict a pile-up of smaller planets at those
radii {\em as long as} the actual type I migration rate of the disruption
remnants is far smaller than the theoretical type I migration rate
\citep[cf.][]{IdaLin08}. Such a pile-up may be testable with the current
exoplanet data.

\section{Acknowledgments}

Theoretical astrophysics research in Leicester is supported by an STFC Rolling
Grant. Andrew King is thanked for discussions. Richard Alexander is thanked
for useful comments on the draft of the manuscript.


\label{lastpage}

\end{document}